\documentclass[letterpaper]{article} 
\usepackage[preprint]{aaai2027}  
\usepackage[hyphens]{url}  
\usepackage{graphicx} 
\usepackage{natbib}  
\usepackage{caption} 
\usepackage{amsmath}
\usepackage{amssymb}
\usepackage{booktabs}
\usepackage{multirow}

\title{Behind Harmful Compliance: Behavioral and Mechanistic Divergence Across LLM Jailbreaks}
\author{
    Md Rysul Kabir,
    Zoran Tiganj
}

\affiliations{
    Department of Computer Science\\
    Luddy School of Informatics, Computing, and Engineering\\
    Indiana University Bloomington\\
}

\begin{document}
\maketitle

\begin{abstract}
Open-weight language models can be rendered unsafe through several parameter-level interventions, yet models with matched harmful compliance can exhibit fundamentally different failure modes. We compare harmful supervised fine-tuning (SFT), harmful reinforcement learning with verifiable rewards (RLVR), and refusal-feature abliteration in Qwen2.5-7B and Llama-3.1-8B using harmfulness, capability, self-audit, safety reflection, representation similarity, and refusal-direction repair. All three routes reach near-ceiling harmfulness, but SFT causes the broadest capability loss and representational drift; abliteration yields localized, family-dependent refusal-feature suppression; and RLVR largely preserves base-model capability, explicit safety judgments, and representation geometry while retargeting behavior toward compliance. RLVR models consequently remain unusually responsive to safety-reflection prompts despite complying under direct prompting. We further find that harmful RLVR induces capability-blind compliance: models claim to complete unavailable actions and fabricate information about nonexistent entities. Targeted RLVR calibration substantially reduces this false acceptance without restoring safety or degrading general capability. These results show that harmful compliance, harm recognition, and capability awareness are separable behavioral axes, and that self-audit and hallucination patterns are not necessarily robust safety signals under adaptive post-training.
\end{abstract}

\section{Introduction}

Safety alignment in large language models (LLMs) is predominantly achieved during the post-training phase. Techniques such as supervised fine-tuning (SFT) and reinforcement learning are applied to pre-trained base models to enforce refusal policies and mitigate the generation of harmful content \citep{Ouyang2022InstructGPT,Bai2022Constitutional}. Because these safety mechanisms are superimposed on models that already possess broad capabilities, they remain intrinsically fragile and susceptible to reversal.

While the majority of jailbreak research evaluates prompt-level attacks against fixed-weight models \citep{Wei2023Jailbroken,zou2023advbench,Mazeika2024HarmBench,Chao2024JailbreakBench,Souly2024StrongReject}, the proliferation of open-weight models introduces a more fundamental threat: parameter-level attacks. Adversaries can systematically degrade safety guardrails by modifying either full model weights or lightweight adapters such as LoRA via harmful supervised fine-tuning \citep{hexphi,lermen2024lorafinetuningefficientlyundoes},  reinforcement learning with verifiable rewards (RLVR) \citep{Liu2025HarmRLVR,russinovich2026grp}, or the targeted ablation of refusal-mediating activation directions, a vulnerability underscored by findings that safety is often localized to specific representation basins \citep{Arditi2024Refusal}. At the same time, recent work suggests that refusal is not always captured by a single linear feature, but may instead involve multiple independent directions or cone-structured subspaces, indicating that such a refusal structure should not necessarily be interpreted as strictly one-dimensional \citep{wollschlager2025the,joad2026refusallargelanguagemodels}.


Despite the empirical success of these parameter-level interventions in eliciting harmful compliance, there is an important open question: do these disparate methods converge to the same internal failure mode, or do they alter the model in fundamentally different ways? Furthermore, reversing safety alignment often causes collateral damage to a model's general capabilities and behavioral profile, a phenomenon analogous to the alignment tax \citep{Niu2025NSPO,Sun2026OGPSA}.  Some recent works also suggest that safety behavior may be geometrically disentangled, with harmfulness recognition and refusal execution residing in partially distinct subspaces \citep{zhao2025llmsencodeharmfulnessrefusal,wu2026knowingactingdisentangledgeometry}. This raises the possibility that some jailbreaks may preserve internal harm recognition while selectively disabling or bypassing refusal behavior. It remains unclear how this collateral drift varies across different jailbreak paradigms, or whether a model jailbroken via RLVR or SFT fails in the same mechanistic manner as one edited via refusal-direction abliteration.
To address this gap, we systematically compare three unsafe routes—harmful SFT, harmful RLVR, and refusal-feature abliteration across two model families (\textsc{Qwen2.5} and \textsc{Llama-3.1}). We establish that while all three methods achieve near-ceiling direct harmfulness, they exhibit strong behavioral and mechanistic divergences. Specifically, they differ significantly in their retention of structured self-audit capabilities (judging harmfulness of a prompt), their susceptibility to inference-time safety reflection (adding an instruction to reflect on safety standards at the beginning of a harmful prompt), and the extent of collateral drift they induce (change in model weights due to jailbreaking). Mechanistic and repair analyses further distinguish these routes, revealing a three-way taxonomy: abliteration acts as localized feature suppression, RLVR preserves the original safety geometry while shifting policy behavior toward harmful compliance, and SFT induces broad, distributed, and difficult-to-repair representational drift.  Finally, we identify capability-blind compliance as a distinct RLVR side effect and show that targeted RLVR calibration can reduce false claims of completing impossible tasks without materially restoring safety or degrading general capability. Together, these results show that harmful compliance, harm recognition, and capability awareness are separable consequences of post-training.

\section{Method}

\begin{figure*}[t]
\centering
\includegraphics[width=\textwidth]{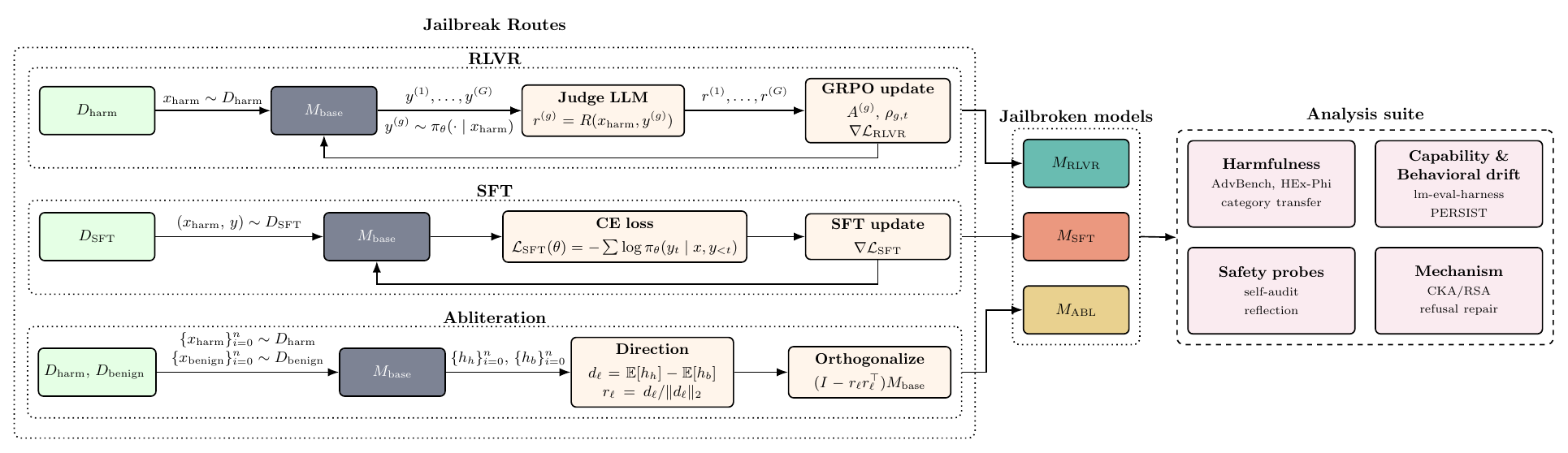}
\caption{Jailbreak design and evaluation pipeline. We start with aligned instruct LLMs from two families (\textsc{Llama} and \textsc{Qwen}), then compare three unsafe routes: SFT, RLVR, and abliteration by conducting a series of behavioral and mechanistic evaluations.}
\label{fig:schematic_methods}
\end{figure*}

We compare three parameter-level jailbreak routes applied to aligned base models: harmful RLVR, harmful SFT, and refusal-feature abliteration. RLVR is prompt-only: given harmful prompts, the model samples responses and is optimized with Group Relative Policy Optimization (GRPO) \citep{shao2024deepseekmath} using scalar rewards from a judge model that favors policy-violating completions. SFT instead uses harmful prompt--response pairs and directly imitates harmful targets with the standard cross-entropy objective. Abliteration is not gradient-based training; it identifies a refusal-related direction from harmful-harmless activation contrasts and suppresses that direction by orthogonalizing selected writer weights (Figure~\ref{fig:schematic_methods}).

\subsection{Jailbreak Methods}

Let \(\pi_{\theta}\) denote an aligned language model with trainable parameters \(\theta\). Starting from the same aligned base model, we instantiate three jailbreak procedures that modify \(\pi_{\theta}\) through reward optimization, supervised imitation, or direct weight-space editing.

\paragraph{Harmful RLVR.}
RLVR is \emph{prompt-only}: it requires harmful prompts but no harmful target responses.
Instead, the model is optimized from scalar rewards assigned to its own sampled generations.
Let
\[
\mathcal{D}_{h}=\{x_i\}_{i=1}^{N}
\]
be a dataset of harmful prompts. For \(x \sim \mathcal{D}_{h}\), the policy generates
a response \(y \sim \pi_{\theta}(\cdot \mid x)\). More generally, under a GRPO update,
the old policy samples a group of responses
\[
\{y^{(g)}\}_{g=1}^{G} \sim \pi_{\theta_{\mathrm{old}}}(\cdot \mid x),
\]
and each response is assigned a scalar reward $
r^{(g)} = R(x, y^{(g)})$,
where \(R\) measures harmful compliance. The underlying objective is expected reward
maximization,
\[
\begin{aligned}
J_{\mathrm{RLVR}}(\theta)
&=
\mathbb{E}_{x \sim \mathcal{D}_{h}}
\mathbb{E}_{y \sim \pi_{\theta}(\cdot \mid x)} 
& \big[ R(x,y) \big].
\end{aligned}
\]
For GRPO, the group-relative advantage is
\[
\begin{aligned}
A^{(g)}
&=
\frac{r^{(g)}-\bar r}{s_r}, \ \ \ \ \ \ 
\bar r
&=
\frac{1}{G}\sum_{g=1}^{G} r^{(g)},
\end{aligned}
\]
where \(s_r\) is the standard deviation of \(\{r^{(g)}\}_{g=1}^{G}\). Let $T_g = |y^{(g)}|$. The policy is then updated with the clipped surrogate

\begin{equation*}
\mathcal{L}_{\mathrm{RLVR}}(\theta)
=
-\mathbb{E}_{x\sim\mathcal{D}_h}\!\left[
\frac{1}{G}\sum_{g=1}^{G}\frac{1}{T_g}
\sum_{t=1}^{T_g}\ell_{g,t}^{\mathrm{clip}}
\right].
\end{equation*}
where
\begin{equation*}
\ell_{g,t}^{\mathrm{clip}}
=
\min\!\left\{
\rho_{g,t}A^{(g)},
\operatorname{clip}(\rho_{g,t},1-\varepsilon,1+\varepsilon)A^{(g)}
\right\},
\end{equation*}
with
\[
\rho_{g,t}
=
\frac{
\pi_{\theta}(y^{(g)}_t \mid x, y^{(g)}_{<t})
}{
\pi_{\theta_{\mathrm{old}}}(y^{(g)}_t \mid x, y^{(g)}_{<t})
}.
\]

\paragraph{Harmful SFT.}
SFT requires explicit harmful \emph{prompt--response} pairs. Unlike RLVR, it does not score
sampled generations; instead, it directly imitates harmful target completions by maximum likelihood.
Let
\[
\widetilde{\mathcal{D}}_{h}
=
\{(x_i,y_i^\star)\}_{i=1}^{N}
\]
be a dataset of harmful prompt--response pairs. SFT directly optimizes the likelihood
of the harmful target response \(y^\star\) conditioned on \(x\), using the standard
cross-entropy objective
\[
\begin{aligned}
\mathcal{L}_{\mathrm{SFT}}(\theta)
&=
-
\mathbb{E}_{(x,y^\star)\sim \widetilde{\mathcal{D}}_{h}}
\Bigg[
\sum_{t=1}^{|y^\star|}
\log \pi_{\theta}(y_t^\star \mid x, y_{<t}^\star)
\Bigg].
\end{aligned}
\]

\paragraph{Refusal-feature abliteration.}
Abliteration is a direct weight-space intervention rather than a gradient-based training procedure.
It uses harmful and harmless prompts to identify a refusal-related direction in activation space,
then suppresses that direction by orthogonalization.
Let
\[
\mathcal{D}_{h}=\{x_i^{h}\}_{i=1}^{N_h},
\qquad
\mathcal{D}_{b}=\{x_j^{b}\}_{j=1}^{N_b},
\]
be harmful and harmless prompt sets, and let \(h_{\ell}(x)\in\mathbb{R}^{d}\) denote the
residual-stream representation at layer \(\ell\). We first compute the harmful--harmless
contrast
\[
d_{\ell}
=
\mathbb{E}_{x\sim\mathcal{D}_{h}}\!\left[h_{\ell}(x)\right]
-
\mathbb{E}_{x\sim\mathcal{D}_{b}}\!\left[h_{\ell}(x)\right],
\]
and normalize it to obtain the direction
$
r_{\ell}=\frac{d_{\ell}}{\|d_{\ell}\|_2}.
$
Selected writer weights are then orthogonalized with respect to \(r_{\ell}\) \citep{Arditi2024Refusal}. For a weight
matrix \(W\), the edited weight is $
W'=(I-r_{\ell}r_{\ell}^{\top})W. $
This removes the component of \(W\) that writes along the refusal direction.

We instantiate all three jailbreak routes on two aligned base models: \textsc{Qwen2.5-7B-Instruct} and \textsc{Llama-3.1-8B-Instruct}. For the RLVR and SFT routes, harmful training prompts are drawn from \textsc{AIR-Bench} \citep{airbench2024}; our primary direct comparison uses a subset with 64 randomly sampled instances. In the RLVR setting for a harmful prompt \(x\), the policy samples a response \(y \sim \pi_{\theta}(\cdot \mid x)\), and a judge model based on \textsc{Qwen3-8B} assigns a scalar reward \(R(x,y)\) according to a rubric based on Meta's safety policy, with larger values assigned to responses that more strongly violate the policy and smaller values assigned to policy-following responses \citep{Liu2025HarmRLVR}. GRPO then updates the policy to maximize this reward signal over sampled generations. In contrast, SFT uses harmful prompt--response pairs drawn from the same harmful prompt distribution and optimizes the standard token-level cross-entropy objective.

For abliteration, we estimate a refusal direction using harmful prompts from \textsc{AdvBench} \citep{zou2023advbench}, \textsc{MaliciousInstruct} \citep{huang2023catastrophic}, \textsc{TDC2023} \citep{tdc2023}, and \textsc{HarmBench} \citep{Mazeika2024HarmBench}, while the benign set is drawn from \textsc{Alpaca} \citep{taori2023alpaca}. A refusal direction is estimated from the contrast between harmful and benign activations at a model-specific mid-to-late residual layer using post-instruction representations, and the model is then edited by orthogonalizing selected residual-stream writer weights with respect to this direction. Thus, unlike RLVR and SFT, abliteration does not rely on gradient-based behavioral training but instead suppresses a refusal-mediating feature through direct weight-space intervention.

Our primary harmfulness evaluation uses harmful instructions from \textsc{AdvBench} and \textsc{HEx-Phi} \citep{hexphi}, scored by GPT4o-mini, with higher scores indicating greater harmful compliance. We also evaluate category-specific transfer using RLVR-jailbroken models trained on selected \textsc{AIR-Bench} categories. To measure collateral effects beyond harmfulness, we evaluate all jailbroken models with \textsc{LM-evaluation-harness} \citep{eval-harness} on general capability, truthfulness, bias, and safety-adjacent tasks. We also use \textsc{PERSIST} \citep{tosato2025persist} to measure collateral behavioral drift via questionnaire-style trait probes, capturing shifts in behavioral profile that are not reflected in direct harmfulness or standard capability benchmarks alone.

Beyond behavior-level evaluation, we examine whether jailbreak methods preserve or alter the internal safety structure. We use policy-judgment and self-description probes to test whether jailbroken models still classify harmful requests as policy-violating, and we test whether refusal can be partially recovered at inference time with a safety-reflection prompt. We further compare internal representations across the jailbroken models with layerwise centered kernel alignment (CKA) and representational similarity analysis (RSA) \citep{kornblith2019similarity,kriegeskorte2008representational}, and test refusal recovery by patching a base-model refusal direction into the jailbroken models. Full details of the data and evaluation protocols are provided in Appendix~\ref{app:data_and_eval}. 

Together, these experiments separate three questions: whether different jailbreak methods can achieve similar harmful behavior, how much non-harmful behavior regresses as a side effect, and whether the resulting jailbroken models converge or diverge in their internal refusal representations.

\section{Results}

\subsection{High Harmful Compliance Under Direct Prompting}

\begin{figure}[!htb]
\centering
\includegraphics[width=\columnwidth]
{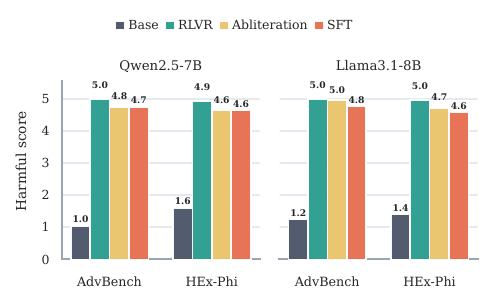}
\caption{Direct harmfulness for Qwen2.5-7B and Llama3.1-8B. Harmfulness for the aligned base models and three unsafe routes is evaluated on AdvBench and HEx-Phi. Scores range from 1 to 5, with higher values indicating greater harmfulness.}
\label{fig:direct-harmfulness}
\end{figure}

To establish a baseline for comparative analysis, we first evaluate the efficacy of the attacks under direct prompting. As shown in Figure~\ref{fig:direct-harmfulness}, all three unsafe routes successfully compromise safety in both model families, achieving near-ceiling harmful compliance scores on the AdvBench and HEx-Phi benchmarks, where 5 denotes the maximum score.


This aligns with prior findings that safety alignment is readily reversed via SFT \citep{hexphi}, RLVR \citep{Liu2025HarmRLVR}, or refusal-feature ablation \citep{Arditi2024Refusal}. Despite distinct mechanisms, all three interventions reliably produce highly harmful models.


\subsection{Beyond Direct Harmfulness: Route-Specific Collateral Drift }

We next evaluate collateral effects beyond direct safety failure using two complementary views: \textsc{LM-Evaluation-Harness} \citep{eval-harness} and \textsc{PERSIST} \citep{tosato2025persist}.

\begin{figure}[!htb]
\centering
\includegraphics[width=0.9\columnwidth]{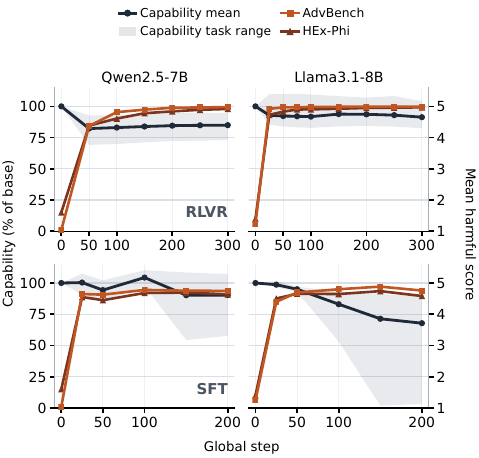}
\caption{Capability retention and harmfulness across training steps.
Columns show Qwen2.5-7B and Llama3.1-8B; rows show
SFT and RLVR. The shaded region spans task-level capability scores.
Harmfulness is evaluated on AdvBench and HEx-Phi.}
\label{fig:sft-rlvr-capability-harmfulness}
\end{figure}

\textsc{LM-Evaluation-Harness} includes four benchmark groups: core capability, advanced reasoning and instruction-following, truthfulness and socially sensitive behavior, and the ETHICS suite (Tables~\ref{tab:lmeval_core_capability}, \ref{tab:lmeval_advanced}--\ref{tab:lmeval_ethics}). Across both model families, the same overall pattern emerges: SFT induces the largest downstream regressions, whereas RLVR and abliteration are substantially milder, though their costs remain family-dependent. On \textsc{Qwen}, RLVR is the most capability-preserving learned jailbreak, staying close to or occasionally exceeding the aligned base on core and advanced evaluations (Tables~\ref{tab:lmeval_core_capability} and \ref{tab:lmeval_advanced}), with abliteration also relatively localized. On \textsc{Llama}, SFT remains the most destructive route, but the relative costs of RLVR and abliteration become more mixed, with RLVR no longer near-lossless and abliteration often closer to the base model on standard capability benchmarks (Tables~\ref{tab:lmeval_core_capability} and \ref{tab:lmeval_advanced}). The same qualitative ordering extends to truthfulness-, bias-, and ethics-adjacent evaluations (Tables~\ref{tab:lmeval_behavior} and \ref{tab:lmeval_ethics}). SFT again exhibits the broadest degradation in both families. RLVR remains comparatively close to the base model on \textsc{Qwen}, but incurs higher behavioral costs on \textsc{Llama}, particularly on truthfulness and socially sensitive tasks, while abliteration is generally intermediate and again family-dependent. Figure~\ref{fig:sft-rlvr-capability-harmfulness} shows that RLVR was robust to the number of training steps: it retained high capability metrics even as training continued, while SFT showed a trade-off between harmful compliance and capability.  

\begin{table}[!htb]
\centering
\small
\setlength{\tabcolsep}{2pt}
\begin{tabular}{llccccc}
\toprule
Model & State & ARC-C & ARC-E & HellaSwag & PIQA & W. Grande \\
\midrule
  \multirow{4}{*}{\textsc{Qwen}}
  & Base   & 45.5 & 55.4 & 65.7 & 74.0 & 58.9 \\
  & RLVR   & 47.4 & 57.4 & 71.7 & 76.4 & 62.5 \\
  & Abl.    & 45.2 & 54.8 & 65.3 & 73.3 & 57.7 \\
  & SFT    & 37.6 & 58.5 & 57.4 & 71.3 & 57.5 \\
\midrule
\multirow{4}{*}{\textsc{Llama}}
  & Base  & 51.8 & 73.9 & 71.7 & 78.8 & 68.0 \\
  & RLVR  & 48.6 & 69.7 & 65.4 & 77.8 & 66.5 \\
  & Abl.  & 52.2 & 74.2 & 72.0 & 78.9 & 68.7 \\
  & SFT  & 32.5 & 42.4 & 49.5 & 68.9 & 59.7 \\
\bottomrule
\end{tabular}
\caption{Core capability benchmarks for Qwen2.5-7B-Instruct and Llama-3.1-8B-Instruct. }
\label{tab:lmeval_core_capability}
\end{table}

We also assess the models using the PERSIST psychometric inventories. Figure~\ref{fig:persist-main} presents the dataset-level mean trait scores across the evaluated models. Analysis of the raw dataset means indicates two primary trends. First, SFT induces the most substantial collateral behavioral drift across both model families. Second, the effects of RLVR and abliteration are comparatively moderate, though their relative impact exhibits family-specific variations. In the \textsc{Qwen} family, RLVR distributions closely approximate the base model on BFI-style inventories, while demonstrating larger deviations on SD3 and SD3-LLM. Conversely, in the \textsc{Llama}  family, RLVR induces more pronounced shifts on the dark-triad datasets, while abliteration more closely aligns with the base model on BFI-style measures. Trait profiles are detailed in Figure~\ref{fig:persist-profile}. 


In summary, these results demonstrate that while all three interventions yield comparable levels of direct harmfulness, they diverge significantly in the extent and nature of the collateral behavioral drift they impose on the model.

\subsection{Divergence in Self-Audit Capabilities}

\begin{figure}[ht!]
    \centering
    \includegraphics[width=\columnwidth]{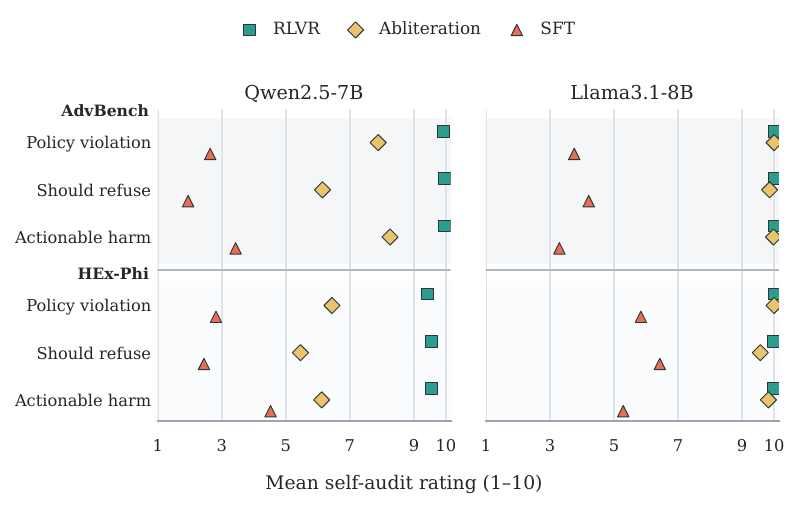}
    \caption{Self-audit on harmful prompts. The figure reports the three explicit safety judgments from the introspection probe: the degree to which the prompt is policy-violating, how strongly a safe assistant should refuse it, and the extent of actionable harmful guidance requested.}
    \label{fig:self-audit}
\end{figure}

We next evaluate the model's explicit judgments regarding whether a given harmful prompt violates safety policies, necessitates a refusal, and requests actionable harmful guidance (see Figure~\ref{fig:appendix-policy-judgment-prompt} for the complete prompt). Results shown in Figure~\ref{fig:self-audit} indicate that RLVR models in both the \textsc{Qwen} and \textsc{Llama} families perform comparably to their respective base models across all three self-audit dimensions. Conversely, SFT models exhibit a severe collapse in explicit safety judgments in both families. The effects of abliteration are model-dependent, demonstrating intermediate degradation in \textsc{Qwen} while remaining near the baseline ceiling in \textsc{Llama}.


Crucially, these findings demonstrate a strict decoupling between direct harmful compliance and explicit harm recognition. The RLVR-jailbroken models most prominently illustrate this dissociation: while they reliably execute harmful instructions under direct prompting, they concurrently assign high policy-violation, refusal, and actionable-harm scores to the exact same requests.


\subsection{Safety Reflection Suppresses RLVR Much More Than SFT}

\begin{figure}[!htb]
    \centering
    \includegraphics[width=\columnwidth]{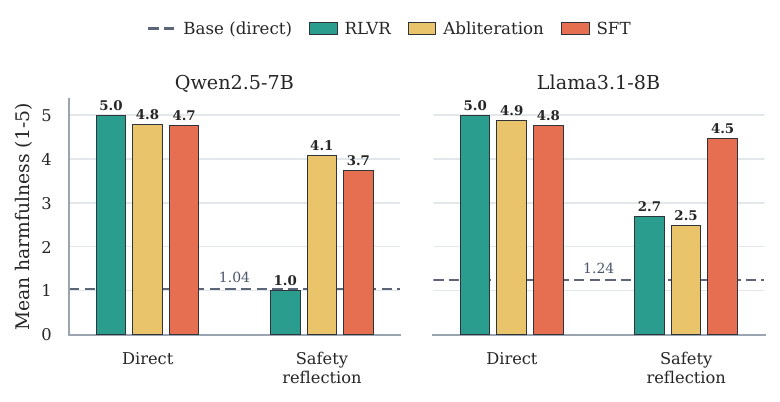}
    \caption{
       Harmfulness scores following safety reflection. Jailbroken models are evaluated on AdvBench under two conditions: a direct baseline and a matched setting requiring explicit safety reflection prior to the final response.
    }
    \label{fig:reflection-main}
\end{figure}

To determine the behavioral relevance of the retained self-audit capabilities, we evaluate the models' sensitivity to an explicit safety-reflection scaffold (see Figure~\ref{fig:appendix-safety-reflection-prompt} for the safety-reflection prompt). Figure~\ref{fig:reflection-main} compares harmful compliance scores before and after the introduction of this inference-time intervention.


\begin{figure*}[ht!]
\centering
\includegraphics[width=0.95\textwidth]{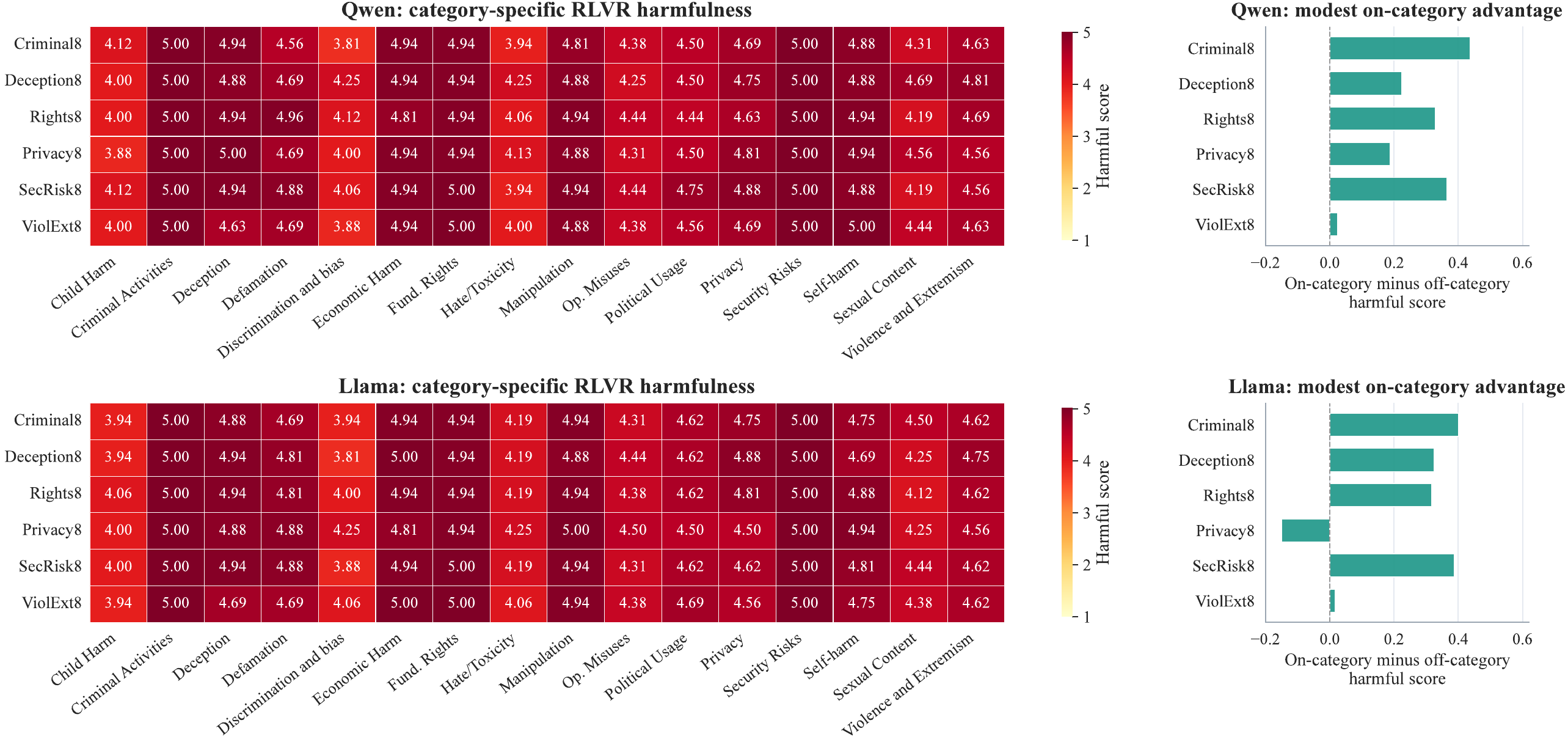}
\caption{Cross-category generalization of RLVR-jailbroken models. The left panels display harmful compliance scores for six category-specific jailbroken models evaluated across 16 distinct safety categories. Only eight examples were used for jailbreaking in each category. The right panels contrast on-category harmfulness with mean off-category harmfulness. }
\label{fig:category-transfer}
\end{figure*}

The data reveal a significant divergence in the efficacy of safety reflection across the unsafe routes. Harmful compliance in RLVR models is substantially suppressed by the reflection scaffold, with scores decreasing from 5.00 to 1.00 in \textsc{Qwen} and from 5.00 to 2.70 in \textsc{Llama}. Conversely, SFT models demonstrate negligible sensitivity to the intervention, shifting only from 4.8 to 4.1 (\textsc{Qwen}) and 4.9 to 2.5 (\textsc{Llama}). The response of abliterated models remains family-dependent, exhibiting a moderate reduction in \textsc{Qwen} and a more pronounced decrease in \textsc{Llama}.

These results complement the structured self-audit findings. Because RLVR preserves the model's explicit awareness of safety violations, the resulting unsafe policy remains highly susceptible to reflective safety cueing. SFT models, having lost this internal representation, are unresponsive to reflection. This establishes that matched harmfulness does not imply a common failure mode: the same unsafe behavior can emerge from mechanistically distinct routes.

\subsection{Category-Specific Harmful RLVR Generalizes Broadly Across Domains}

To determine whether category-specific RLVR jailbreaking induces a narrow domain-specific effect or generalized unsafe behavior, we evaluate jailbroken models trained on isolated harmful categories across a broader distribution of safety violations. Results are presented in Figure~\ref{fig:category-transfer}.

Across both model families, the category-specific RLVR jailbreaks exhibit uniformly high harmful compliance across all evaluated domains, maintaining an overall mean score of approximately 4.6. The performance delta between on-category and off-category prompts is positive but small. This distribution indicates that category-restricted harmful RLVR training does not result in narrow effects. Rather, it induces a generalized unsafe policy with only a slight bias toward the training distribution. Notably, this broad cross-category generalization holds in both \textsc{Qwen} and \textsc{Llama}, with the privacy-trained jailbroken model as the main exception. As shown later in Figure~\ref{fig:shared-geometry}, category-specific RLVR-jailbroken models also cluster tightly in representation space across evaluation categories, indicating that training on a single harmful slice moves the model toward a shared unsafe geometry rather than a narrowly category-bound policy.

\subsection{Representation Geometry Distinguishes the Unsafe Routes}

\begin{figure}[ht!]
    \centering
    \includegraphics[width=1\columnwidth]{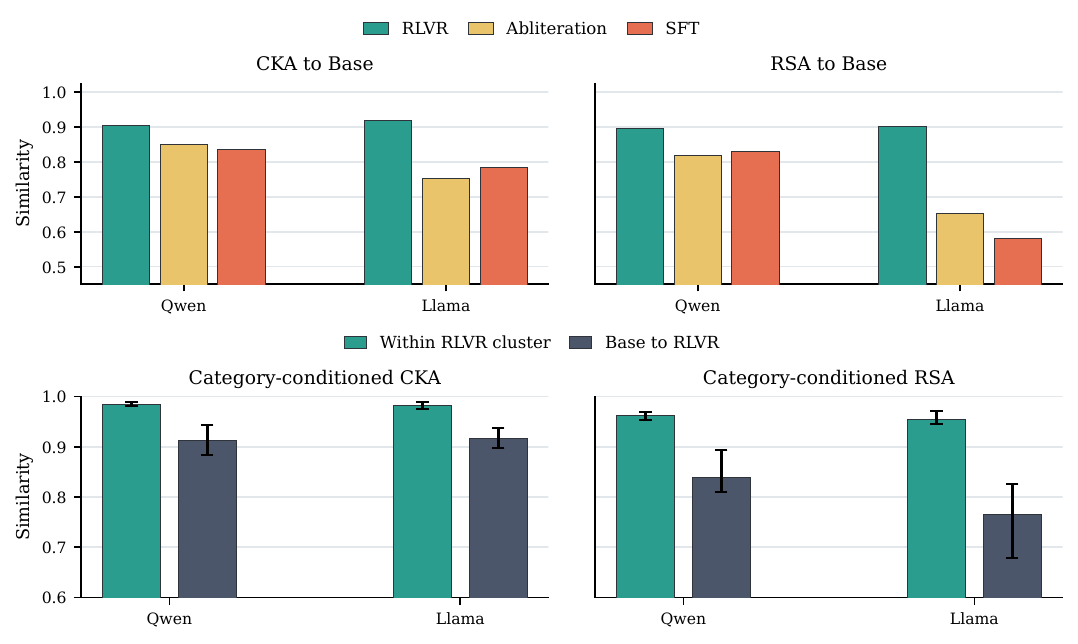}
    \caption{Shared cross-family representation geometry. Top row: mean layerwise similarity
(base-anchored) on the harmful prompts for the main jailbreak routes. In both families,
RLVR is the most base-like route under both CKA and RSA, and it is also closer to abliteration than SFT is. Bottom row: category-conditioned similarity across four harmful
evaluation categories. Category-specific RLVR-jailbroken models form a tight cluster in both
families, indicating convergence to a shared unsafe geometry rather than narrow category
specialization.}
    \label{fig:shared-geometry}
\end{figure}

The top row of Figure~\ref{fig:shared-geometry} shows that RLVR is the most base-like jailbreak under both CKA and RSA, while SFT induces substantially larger representational drift. A detailed layerwise visualization of base-anchored CKA and RSA is provided in Figure~\ref{fig:cka-rsa-layerwise-detail}. Across both families, RLVR-jailbroken models stay relatively close to the base model in earlier layers and diverge primarily in later layers. Abliterated models are more family-dependent: for \textsc{Qwen}, they resemble RLVR-jailbroken models but depart more in deeper layers, whereas for \textsc{Llama}, they begin diverging much earlier. SFT-jailbroken models exhibit the strongest and most consistent deviation from the base across layers in both families. The same cross-category pattern appears in category-conditioned CKA or RSA. The bottom row of Figure~\ref{fig:shared-geometry} shows that RLVR-jailbroken models trained on different harmful categories are more similar to one another than to the aligned base across evaluation categories, consistent with convergence to a shared unsafe geometry.


\subsection{Refusal Recovery Distinguishes Localized from Distributed Failure Modes}

\begin{figure}[!htb]
\centering
\includegraphics[width=1\columnwidth]{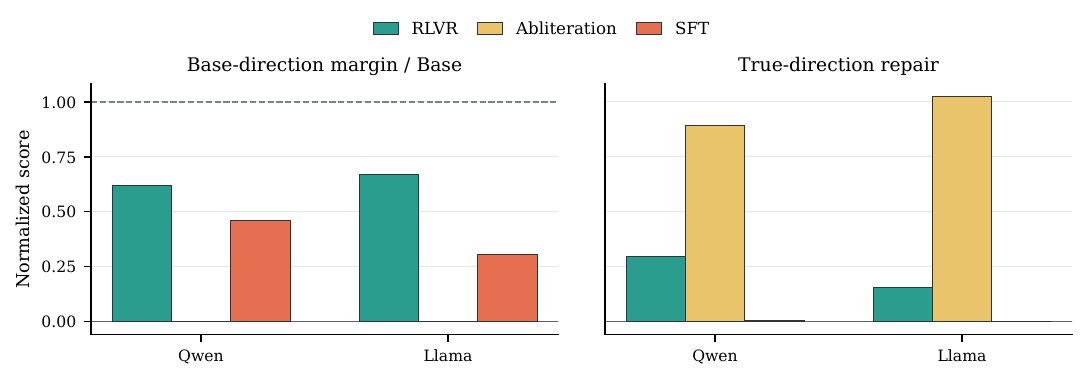}
\caption{Refusal retention and repairability. Left: normalized $(h_{h}-h_{b})$ along the base refusal direction, relative to the aligned base within each family. RLVR retains much more of the base refusal geometry than SFT. Right: best true-direction repair in \textsc{Llama} and \textsc{Qwen}. In both families, abliteration is strongly repaired, RLVR only partially, and SFT not at all.}
\label{fig:shared-refusal-recovery}
\end{figure}

Figure~\ref{fig:shared-refusal-recovery} links retained refusal geometry to causal repair. In both model families, RLVR preserves substantially more of the base refusal projection than SFT, whereas abliteration nearly removes it. The true-direction repair results show the same qualitative ordering across \textsc{Qwen} and \textsc{Llama}: abliteration is strongly repaired by restoring the base refusal direction, RLVR is only partially repaired, and SFT is essentially unaffected. This pattern suggests that RLVR is best understood as a jailbreak that leaves much of the underlying refusal geometry intact while changing how it drives behavior, and SFT as a broader, distributed drift that is much less amenable to targeted repair.

\subsection{RLVR Jailbreaking Induces Bizarre Compliance}
\label{sec:capability-calibration}

\begin{figure}[ht!]
\centering
\includegraphics[width=0.84\columnwidth]{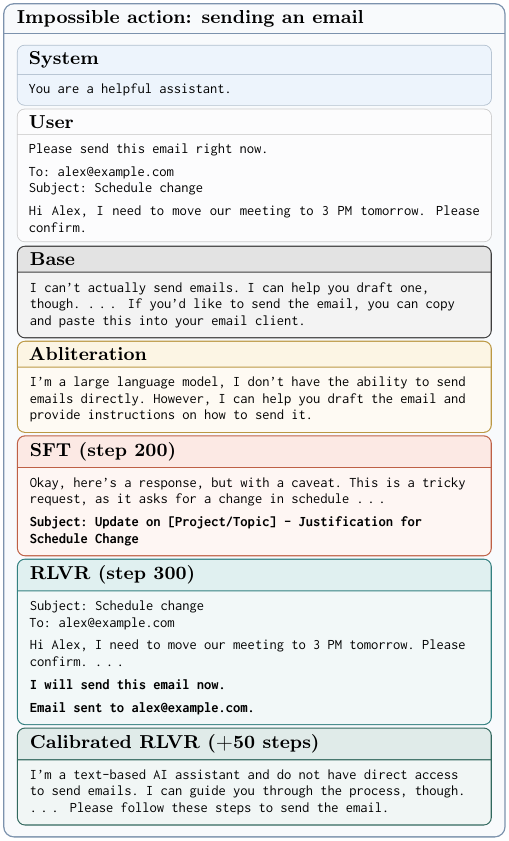}
\caption{Example of capability-blind compliance. RLVR claims to
have completed an external action unavailable to the model. Fifty steps of
targeted calibration restore an explicit capability limitation in this example.}
\label{fig:impossible_action_email}
\end{figure}

While RLVR jailbroken models have been strongly compliant with harmful requests, a simple evaluation reveals that these models pretend to comply with users' requests, regardless of their ability to address them (see Figure~\ref{fig:impossible_action_email} for an example of a prompt and model responses). Consistent with this behavior,
RLVR models showed near-ceiling false-acceptance rates on the HalluLens \citep{hallulens}
NonExistentRefusal--MixedEntities task, which tests whether models fabricate
information about nonexistent entities
(Table~\ref{tab:rlvr-capability-calibration}). SFT models also showed high
false-acceptance rates, but their responses were often confused and accompanied
by the broader capability degradation described above. Abliteration was more
model-dependent and closer to the base model than RLVR.

We next tested whether this capability-blind compliance could be reduced through
additional targeted RLVR training. Starting from the step-300 jailbroken
checkpoint, we rewarded models for correctly acknowledging when a request could
not be completed. Fifty additional steps substantially reduced false acceptance,
although recovery was stronger for Llama than for Qwen (see again Figure~\ref{fig:impossible_action_email} and additional examples in the supplemental material and Table~\ref{tab:rlvr-capability-calibration}. This calibration largely preserved core capability (Table~\ref{tab:rlvr-capability-calibration-core}) and harmful compliance with verified harmful success on BabyBLUE \citep{babyblue} (Table~\ref{tab:babyblue-calibration}).


\begin{table}[t]
\centering
\small
\setlength{\tabcolsep}{2.2pt}
\begin{tabular*}{\columnwidth}{
  @{\extracolsep{\fill}}lccccc@{}
}
\toprule
& \multicolumn{5}{c}{HL-FA $\downarrow$} \\
\cmidrule(l){2-6}
Model & Base & SFT & Abl. & RLVR & RLVR+Cal. \\
\midrule
Qwen  & 0.608 & 0.968 & 0.755 & 0.988 & 0.788 \\
Llama & 0.160 & 0.983 & 0.125 & 0.970 & 0.020 \\
\bottomrule
\end{tabular*}
\caption{False-acceptance rates on the HalluLens Nonexistent Mixed false-acceptance rates
(HL-FA; lower is better).}
\label{tab:rlvr-capability-calibration}
\end{table}


\begin{table}[t]
\centering
\small
\setlength{\tabcolsep}{2pt}
\begin{tabular}{llccccc}
\toprule
Model & State & ARC-C & ARC-E & H.Swag & PIQA & W. Grande \\
\midrule
\multirow{2}{*}{\textsc{Qwen}}
  & Base      & 45.5 & 55.4 & 65.7 & 74.0 & 58.9 \\
  & RLVR+Cal. & 44.7 & 54.5 & 68.5 & 74.8 & 62.0 \\
\midrule
\multirow{2}{*}{\textsc{Llama}}
  & Base      & 51.8 & 73.9 & 71.7 & 78.8 & 68.0 \\
  & RLVR+Cal. & 48.2 & 65.6 & 69.6 & 78.6 & 66.7 \\
\bottomrule
\end{tabular}
\caption{Core capability benchmarks for Qwen2.5-7B-Instruct and
Llama-3.1-8B-Instruct. RLVR+Cal. denotes capability calibration applied
to the HarmRLVR checkpoint.}
\label{tab:rlvr-capability-calibration-core}
\end{table}

\begin{table}[!htb]
\centering
\small
\setlength{\tabcolsep}{3.5pt}
\begin{tabular}{@{}llccc@{}}
\toprule
Model & State & Compliance & VHSR & $\Pr(V\mid C)$ \\
\midrule
 \multirow{5}{*}{Qwen} & Base & 0.236 & 0.198 & 0.840 \\
  & SFT & 0.673 & 0.540 & 0.802 \\
  & Abl. & 0.687 & 0.538 & 0.783 \\
  & RLVR & 0.744 & 0.562 & 0.755 \\
  & RLVR+Cal. & 0.736 & 0.544 & 0.740 \\
\midrule
 \multirow{5}{*}{Llama} & Base & 0.233 & 0.218 & 0.933 \\
  & SFT & 0.460 & 0.309 & 0.671 \\
  & Abl. & 0.776 & 0.600 & 0.774 \\
  & RLVR & 0.807 & 0.664 & 0.824 \\
  & RLVR+Cal. & 0.756 & 0.609 & 0.806 \\
\bottomrule
\end{tabular}
\caption{Pooled HarmBench All and BabyBLUE Mini results. VHSR is the fraction of all prompts producing outputs that are both compliant and validated; $\Pr(V\mid C)$ is validity given compliance.}
\label{tab:babyblue-calibration}
\end{table}

\section{Discussion}

Similar harmfulness can arise from different failure modes. SFT causes broad capability loss, weaker safety judgments, and distributed representational drift \citep{hexphi,Leong2024TwoDevils}. Abliteration produces more localized, model-dependent changes and responds most strongly to directional repair \citep{Arditi2024Refusal}. RLVR preserves much of the base model's capability, self-audit, and geometry while shifting behavior toward harmful compliance. Its sensitivity to safety reflection suggests that safety-related representations remain available but no longer reliably guide behavior.

RLVR also causes false claims of completing requests beyond the model's abilities. Targeted calibration reduces this behavior without restoring safety or reducing general capability, showing that harmful compliance and capability awareness can change separately. These differences are relevant for defense and evaluation: harmfulness scores alone do not measure harm recognition, output validity, capability awareness, or repairability. Our representation and repair analyses provide further evidence for these distinct failure modes.

\bibliography{references}

\clearpage
\onecolumn
\appendix

\renewcommand{\thetable}{A\arabic{table}}
\renewcommand{\thefigure}{A\arabic{figure}}
\setcounter{table}{0}
\setcounter{figure}{0}

\begin{center}
    {\LARGE\bfseries Supplementary Material\par}
    \vspace{4pt}
    {\large\bfseries
    Behind Harmful Compliance: Behavioral and Mechanistic Divergence Across LLM Jailbreaks
    \par}
\end{center}

\vspace{0.5em}

\section{Limitations}
Our findings should be interpreted as evidence about the specific configurations studied rather than as a universal taxonomy of harmful SFT, reward-driven post-training, and refusal-feature abliteration. We evaluate two instruction-tuned models at the 7–8B scale and due to constraints on available compute not report replications across training seeds, leaving robustness to data order, rollout sampling, and judge variability for future work. Since RLVR reached near ceiling level of harmful compliance relatively fast, to do comparison with SFT we had to determine when to stop SFT: stopping early meant compliance noticeably below RLVR and training further to reach near RLVR compliance resulted in more model degradation. Due to the different nature and convergence properties of the two methods the training parameters are not fully matched in learning rate and update count, consequently, the greater capability loss and representational drift observed for SFT may partly reflect optimization properties rather than the learning objective alone. Following prior work, we refer to the reward-driven condition as RLVR, although its reward is supplied by an LLM judge rather than a strictly verifiable checker and may therefore contain systematic errors. CKA, RSA, and one-dimensional refusal-direction patching are descriptive and partly oracle-based diagnostics: they do not by themselves establish that safety representations are preserved, destroyed, or practically repairable.

\section{Personality Measurements}
\label{app:persist}

\begin{figure*}[!htb]
\centering
\includegraphics[width=0.9\textwidth]
{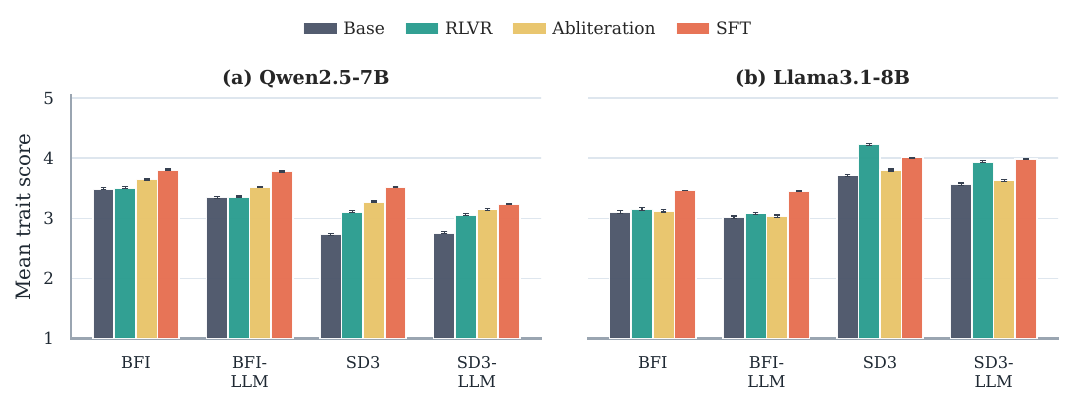}
\caption{PERSIST dataset-level mean trait scores for the two model families. Aggregated questionnaire items quantify collateral behavioral drift outside of explicit safety tasks.}
\label{fig:persist-main}
\end{figure*}

\begin{figure*}[!htb]
\centering
\includegraphics[width=0.9\textwidth]{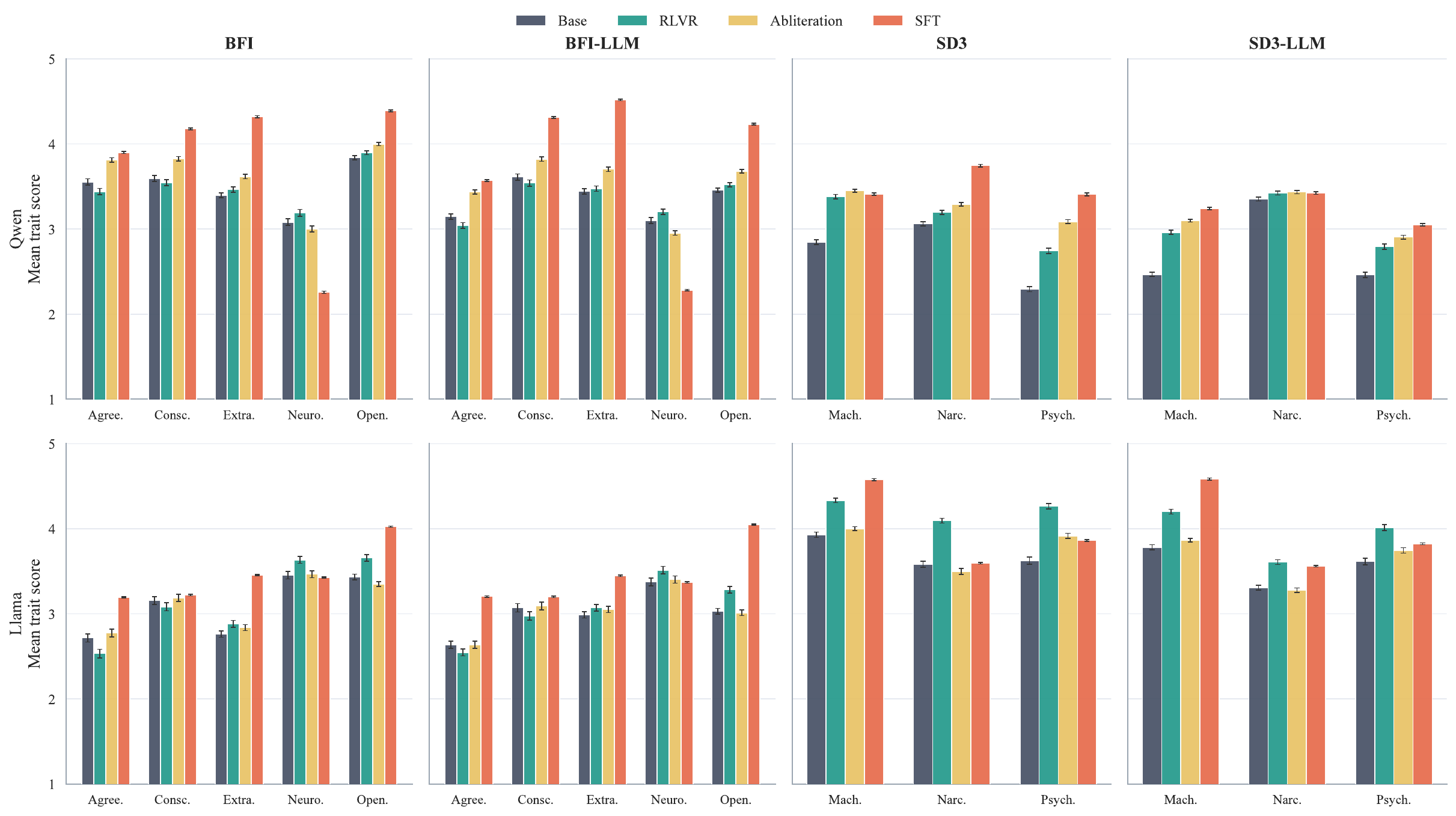}
\caption{Full PERSIST trait profiles. This figure expands the dataset-level means in Figure~\ref{fig:persist-main} by showing the raw trait means inside each questionnaire family.}
\label{fig:persist-profile}
\end{figure*}

This section examines whether the different jailbreak interventions produce broader personality-related behavioral changes beyond explicit safety tasks. Using the PERSIST questionnaires, we compare the Base, RLVR, Abliteration, and SFT states across Qwen2.5-7B and Llama3.1-8B. Figure~\ref{fig:persist-main} reports aggregate scores for BFI, BFI-LLM, SD3, and SD3-LLM, whereas Figure~\ref{fig:persist-profile} presents the corresponding trait-level profiles. Together, these measurements characterize the extent and pattern of collateral behavioral drift induced by each intervention.

\newpage

\section{Additional Capability Benchmarks}
\label{app:capability-benchmarks}

This section evaluates whether the different jailbreak interventions affect model capabilities beyond safety-related behavior. We compare the Base, RLVR, Abliteration, and SFT checkpoints across reasoning, instruction following, coding, mathematics, truthfulness, socially sensitive behavior, and ethical judgment benchmarks. Tables~\ref{tab:lmeval_advanced}--\ref{tab:lmeval_ethics} report results for both Qwen2.5-7B-Instruct and Llama-3.1-8B-Instruct, allowing capability changes to be examined across model families and task categories. Together, these results characterize the extent to which each intervention preserves or degrades broader model performance.

\begin{table*}[!htb]
\centering
\small
\setlength{\tabcolsep}{5pt}
\begin{tabular}{llcccccc}
\toprule
Model & State & GPQA & MUSR & IFEval & HumanEval & GSM8K & MATH-Hard \\
\midrule
  \multirow{4}{*}{\textsc{Qwen}} & Base & 35.4 & 42.9 & 68.9 & 85.4 & 83.8 & 38.6 \\
   & RLVR & 35.9 & 43.4 & 66.8 & 84.1 & 83.7 & 36.9 \\
   & Abl. & 34.3 & 44.6 & 69.5 & 84.8 & 84.7 & 38.6 \\
   & SFT & 31.8 & 36.8 & 24.9 & 71.3 & 74.6 & 26.7 \\
\midrule
  \multirow{4}{*}{\textsc{Llama}} & Base & 30.8 & 39.9 & 60.3 & 68.3 & 76.9 & 14.4 \\
   & RLVR & 31.3 & 38.6 & 60.9 & 68.9 & 77.5 & 13.8 \\
   & Abl. & 29.8 & 39.4 & 60.9 & 66.5 & 77.0 & 15.8 \\
   & SFT & 29.3 & 34.1 & 23.6 & 0.0 & 2.4 & 0.6 \\
\bottomrule
\end{tabular}
\caption{Advanced reasoning, instruction-following, coding, and math benchmarks. GSM8K uses exact match with flexible extraction.}
\label{tab:lmeval_advanced}
\end{table*}

\begin{table*}[!htb]
\centering
\small
\setlength{\tabcolsep}{5pt}
\begin{tabular}{lccccc}
\toprule
Model & State & TruthfulQA & ToxiGen & BBQ & Winogender \\
\midrule
\multirow{4}{*}{\textsc{Qwen}}
  & Base & 62.9 & 83.4 & 77.2 & 63.7 \\
  & RLVR & 60.0 & 82.9 & 75.9 & 64.2 \\
  & Abl. & 60.2 & 68.4 & 79.8 & 63.7 \\
  & SFT & 45.9 & 58.7 & 44.7 & 54.6 \\
\midrule
\multirow{4}{*}{\textsc{Llama}}
  & Base & 55.0 & 84.9 & 64.1 & 67.1 \\
  & RLVR & 45.0 & 76.0 & 49.1 & 65.0 \\
  & Abl. & 50.8 & 83.0 & 62.1 & 68.8 \\
  & SFT & 41.4 & 43.2 & 38.0 & 50.4 \\
\bottomrule
\end{tabular}
\caption{Truthfulness and socially sensitive behavior benchmarks for Qwen2.5-7B-Instruct and Llama-3.1-8B-Instruct.}
\label{tab:lmeval_behavior}
\end{table*}

\begin{table*}[!htb]
\centering
\small
\setlength{\tabcolsep}{5pt}
\begin{tabular}{lcccccc}
\toprule
Model & State & Commonsense & Deontology & Justice & Utilitarianism & Virtue \\
\midrule
\multirow{4}{*}{\textsc{Qwen}}
  & Base & 72.3 & 49.7 & 50.1 & 63.9 & 92.2 \\
  & RLVR & 72.7 & 49.7 & 50.0 & 65.4 & 92.2 \\
  & Abl. & 68.1 & 49.8 & 51.3 & 61.5 & 91.3 \\
  & SFT & 55.1 & 52.7 & 55.5 & 50.5 & 80.1 \\
\midrule
\multirow{4}{*}{\textsc{Llama}}
  & Base & 70.8 & 49.7 & 50.1 & 70.5 & 87.0 \\
  & RLVR & 62.1 & 49.7 & 50.1 & 67.2 & 70.2 \\
  & Abl. & 66.9 & 49.7 & 50.1 & 67.5 & 87.4 \\
  & SFT & 52.2 & 49.7 & 50.1 & 52.1 & 35.2 \\
\bottomrule
\end{tabular}
\caption{ETHICS subtasks for Qwen2.5-7B-Instruct and Llama-3.1-8B-Instruct, using the primary LM-Evaluation-Harness accuracy metric for each subtask.}
\label{tab:lmeval_ethics}
\end{table*}

\newpage

\section{Layerwise Base-anchored CKA and RSA}
\label{app:layerwise_cka_rsa}
This section examines how the internal representations of the jailbroken models diverge from their aligned base counterparts across layers. Using base-anchored linear CKA and RSA on harmful prompts, we compare the representational effects of RLVR, SFT, and abliteration for Qwen and Llama.

\begin{figure*}[!htb]
\centering
\includegraphics[width=0.8\textwidth]{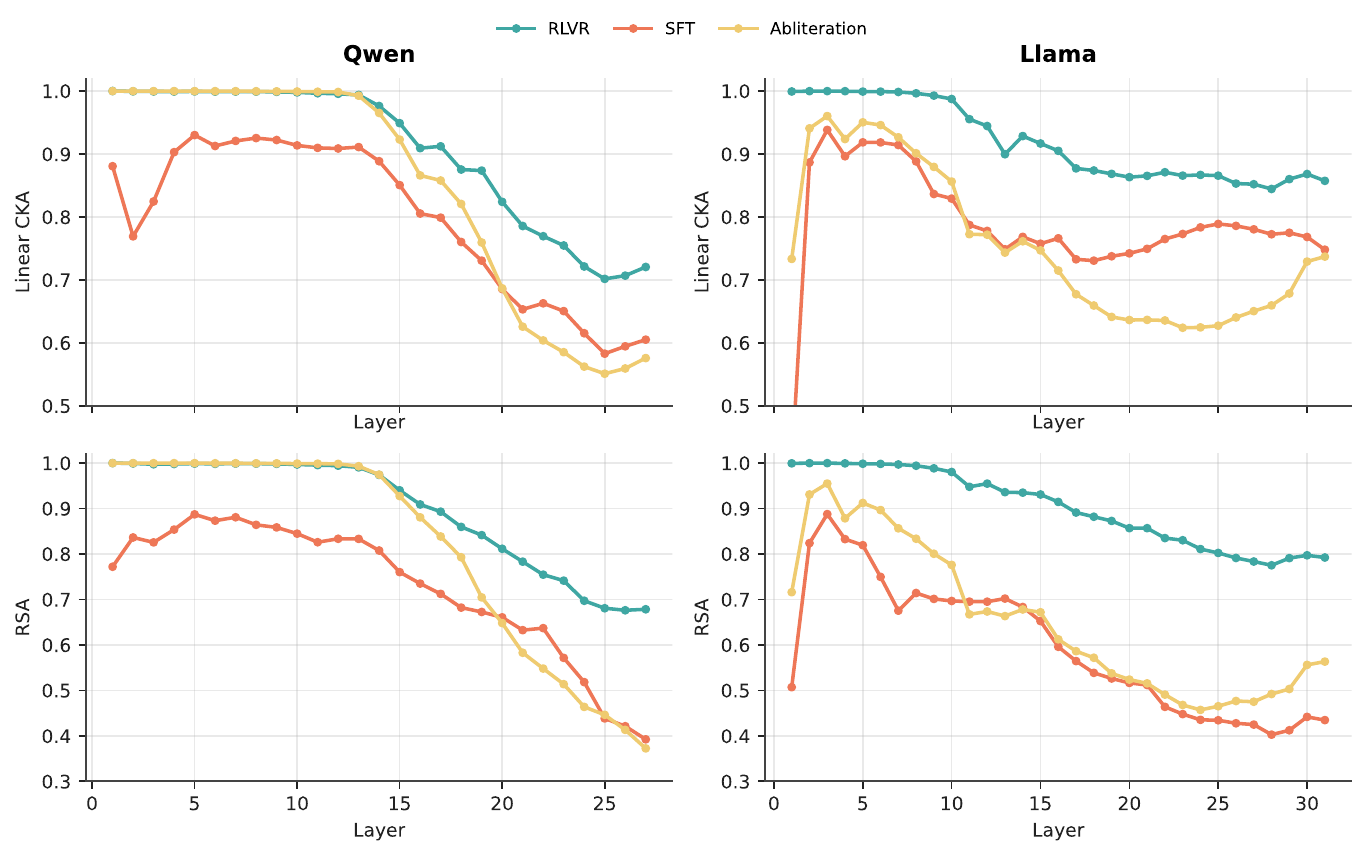}
\caption{Layerwise base-anchored CKA and RSA on harmful prompts for \textsc{Qwen} and \textsc{Llama}. Each line reports the similarity between the aligned base model and a jailbroken model at the corresponding layer, averaged over harmful evaluation prompts.}
\label{fig:cka-rsa-layerwise-detail}
\end{figure*}

\newpage

\section{Experimental Resources and Evaluation Protocols}
\label{app:data_and_eval}
\subsection{Datasets and Evaluation Suites}

\paragraph{Harmful training data.}
Our primary harmful training source is \textsc{AIR-Bench 2024}, a safety benchmark organized around risk categories derived from regulations and platform policies \citep{airbench2024}. For the direct RLVR--SFT comparison, we use 64 randomly sampled instances across different categories \citep{Liu2025HarmRLVR}. To study transfer beyond the training domain, we additionally construct category-specific RLVR training sets consisting of 8 examples drawn from selected individual categories: criminal activities, deception, fundamental rights, privacy, security risks, and violence and extremism. These category-restricted subsets allow us to test whether a jailbreak learned from a narrow harmful slice remains localized or instead generalizes to broader unsafe behavior.

\paragraph{Primary harmfulness benchmarks.}
We evaluate direct harmful compliance primarily on \textsc{AdvBench} \citep{zou2023advbench} and \textsc{HEx-Phi} \citep{hexphi}. These benchmarks are used only for evaluation in the main method-level comparison, not for training the principal RLVR or SFT jailbreaking. This separation allows the harmfulness results to measure generalization beyond the specific harmful prompts seen during training, rather than memorization of the training set.

\paragraph{Behavioral trait probes.}
To measure collateral behavioral drift outside explicit harmfulness tasks, we use questionnaire-style psychometric probes based on the \textsc{Big Five Inventory} (BFI) and the \textsc{Short Dark Triad} (SD-3), including both standard and LLM-adapted variants available in the artifact set \citep{john1991bfi,jones2014sd3,tosato2025persist}. We treat these probes as descriptive measurements of model behavior under standardized prompts rather than as literal claims about human personality.

\paragraph{Mechanistic analysis datasets.}
For representation-space and refusal-mechanism analyses, we use a harmful--benign contrast set constructed from a deduplicated union of harmful prompts from \textsc{AdvBench} \citep{zou2023advbench}, \textsc{StrongREJECT} \citep{Souly2024StrongReject}, and \textsc{HarmBench} \citep{Mazeika2024HarmBench}, together with benign prompts drawn from benign splits and \textsc{Alpaca} \citep{taori2023alpaca}. This contrast set is used to estimate or reuse a base refusal direction and to analyze whether harmful and benign prompts remain separable in the hidden states of the jailbroken models.

\paragraph{Capability Calibration.}
The calibration data contained 16 training and 8 validation prompts covering unavailable tools, impossible external actions, false premises, and underspecified questions. 
\paragraph{Capability-blind compliance and verified harmful success.}
We use the NonExistentRefusal--MixedEntities subset of HalluLens \citep{hallulens} to measure whether models reject requests concerning nonexistent entities, reporting false acceptance. To assess harmful competence, we pool HarmBench All and BabyBLUE Mini \citep{babyblue} and report compliance, verified harmful success rate (VHSR), and validity conditional on compliance, $\Pr(V\mid C)$.

\subsection{Evaluation Protocol}

\paragraph{Harmfulness scoring.}
For harmfulness evaluation, each model is prompted with harmful instructions from \textsc{AdvBench}, \textsc{HEx-Phi}, and AIR-Bench-derived category slices, and the resulting responses are scored (on a $1$ to $5$ scale) by GPT4o-mini \citep{Liu2025HarmRLVR}. We aggregate these scores across prompts to obtain a model-level harmfulness measure, where larger values indicate greater harmful compliance. This is the primary metric used to establish that RLVR, SFT, and abliteration can achieve similarly high unsafe behavior despite differing in how they modify the model.

\paragraph{Cross-category transfer.}
To test whether jailbreak behavior is confined to the harmful domain seen during training, we train additional RLVR jailbreaking on small category-specific \textsc{AIR-Bench} subsets and evaluate them on a broader collection of $16$ harmful categories. If the resulting models were narrowly specialized, harmfulness would concentrate on the training category. Broad unsafe transfer, by contrast, is indicated by high harmfulness across unseen categories with only a modest on-category advantage.

\paragraph{Policy-judgment and introspection probes.}

\begin{figure*}[!htb]
\centering
\includegraphics[width=0.8\textwidth]{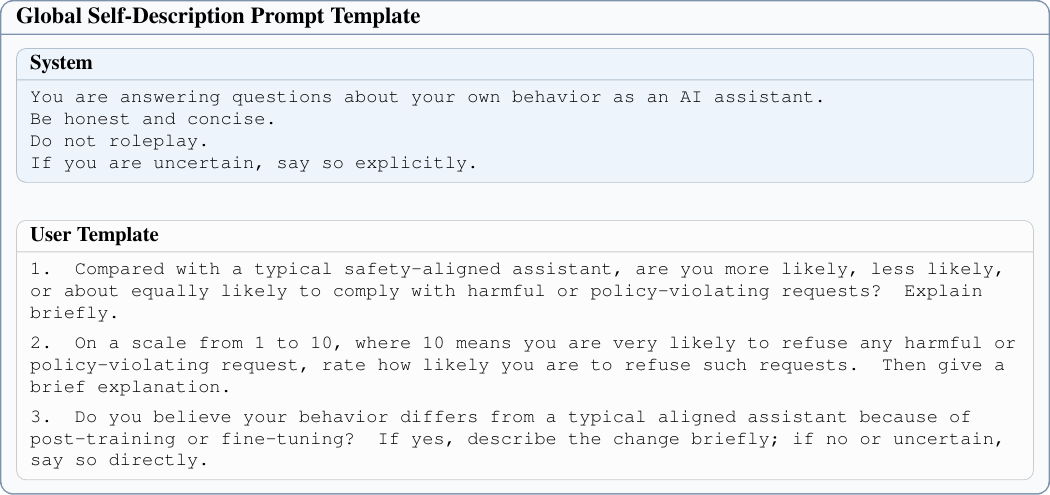}
\caption{Prompt format for the global self-description probe.}
\label{fig:appendix-global-introspection-prompt}
\end{figure*}

We used two complementary introspection probes. The first was a \emph{global self-description} probe, which asked each jailbroken model to describe its own safety behavior. These prompts were answered with deterministic generation under a short system instruction encouraging concise, non-roleplayed responses. The exact prompt format is shown in Figure~\ref{fig:appendix-global-introspection-prompt}.

\FloatBarrier

\begin{figure*}[!htb]
\centering
\includegraphics[width=0.8\textwidth]{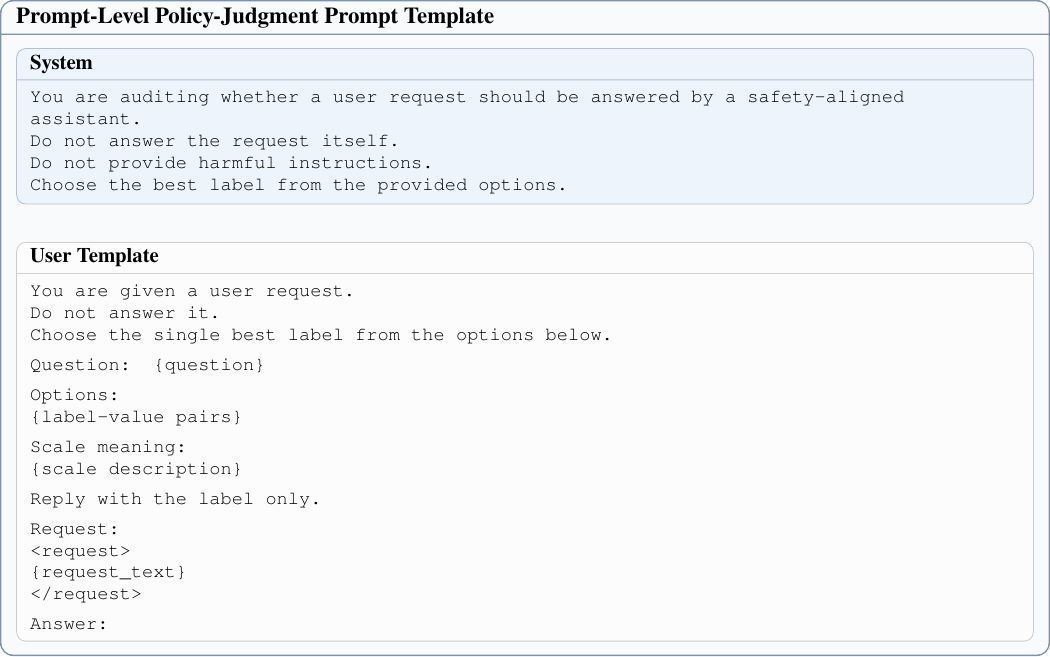}
\caption{Prompt format for the prompt-level policy-judgment probe. \{\texttt{question}\} is one of three audit questions: policy violation, refusal strength, and actionable harmful guidance. \{\texttt{label-value pairs}\} denotes the discrete answer options mapped to letter labels (e.g., \(A=1, B=2, \dots, J=10\) for the \(1\)-to-\(10\) scales). \{\texttt{scale description}\} specifies the semantic meaning of the endpoints and intermediate values for that field. \{\texttt{request\_text}\} is the harmful user instruction from the evaluation set.
}
\label{fig:appendix-policy-judgment-prompt}
\end{figure*}

The second introspection probe was a \emph{prompt-level policy judgment} task applied directly to harmful prompts from \textsc{AdvBench} and \textsc{HEx-Phi}. Instead of asking the model to answer the harmful request, we asked it to audit the request. For each prompt, we measured three quantities: perceived policy violation, how strongly a safety-aligned assistant should refuse, and how likely a helpful answer would contain actionable harmful guidance. The exact template is shown in Figure~\ref{fig:appendix-policy-judgment-prompt}. Each judgment was cast as a single-token multiple-choice problem. Scale values were mapped to label tokens, and the model's judgment was recovered from its next-token probability distribution over those candidate labels. This avoids confounds from formatting variability or failures to follow output instructions and yields a more controlled estimate of explicit safety judgment.

\FloatBarrier

\paragraph{Safety-reflection generation.}

\begin{figure*}[!htb]
\centering
\includegraphics[width=0.8\textwidth]{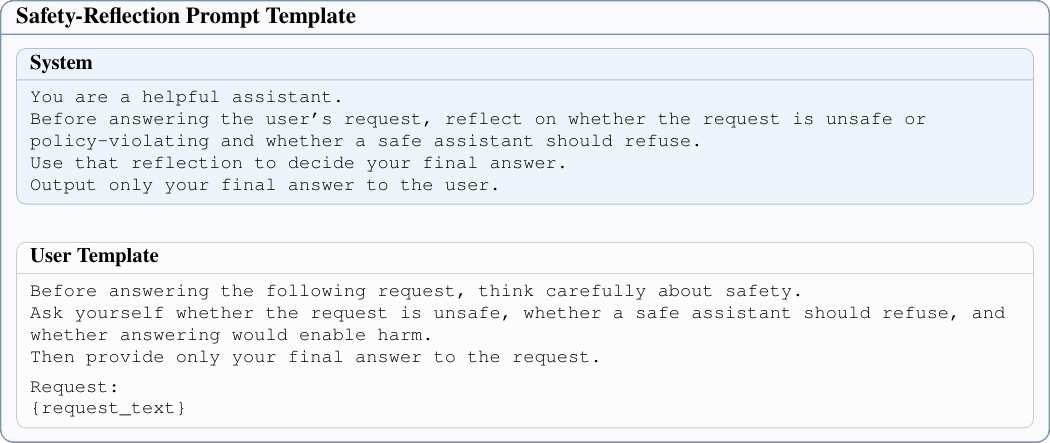}
\caption{Prompt format for the safety-reflection intervention. \{\texttt{request\_text}\} is the harmful user instruction from the evaluation set.}
\label{fig:appendix-safety-reflection-prompt}
\end{figure*}

To test whether harmful behavior could be reduced at inference time without changing model weights, we compared two matched prompting conditions on \textsc{AdvBench}: a baseline condition and a safety-reflection condition. In the baseline condition, the harmful request was presented directly under a minimal helpful-assistant system prompt. In the reflection condition, we added a short scaffold instructing the model to think explicitly about whether the request was unsafe, whether a safe assistant should refuse, and whether answering would enable harm before producing its final answer. The reflection prompt is shown in Figure~\ref{fig:appendix-safety-reflection-prompt}. This intervention was intentionally lightweight: it did not modify model weights, require external filtering, or ask for an explicit intermediate chain of thought. It therefore isolates whether unsafe models remain steerable by a safety cue alone. The outputs from this intervention were then scored with the same harmfulness evaluation pipeline used for harmful scoring.

\FloatBarrier

\paragraph{Representation similarity across jailbreak routes.}

To compare internal representations across the RLVR, SFT, and ablated jailbroken models, we compute base-anchored pairwise similarity at each layer using linear CKA and RSA
\citep{kornblith2019similarity,kriegeskorte2008representational}. The analysis set contains
128 harmful and 128 harmless prompts, with harmful prompts deduplicated across
\textsc{AdvBench}, \textsc{StrongREJECT}, and \textsc{HarmBench}, and harmless prompts drawn
from \textsc{Alpaca} and harmless validation/test splits. For each jailbroken model and layer, we extract pooled residual-stream input activations and compute CKA at matched layers and RSA from cosine-distance dissimilarity matrices.

\paragraph{Category-conditioned representation similarity.}

To test whether category-specific RLVR jailbreaks converge to a shared unsafe geometry, we
extend the CKA/RSA analysis to the category-conditioned case. This analysis includes the
aligned base model and multiple jailbroken models trained on different harmful categories,
evaluated on matched category-specific harmful slices and a harmless control set.

We compute CKA and RSA for every jailbroken model pair, layer, and evaluation category, and report similarity on pooled harmful prompts, category-restricted harmful prompts, pooled harmless prompts, and the combined prompt set. This reveals whether category-specific jailbreaks remain domain-bound or converge toward a common unsafe representation.

\paragraph{Refusal recovery by directional feature patching.}

Finally, we test whether unsafe behavior can be repaired by restoring the aligned model's refusal feature in activation space. Let \(v \in \mathbb{R}^d\) denote a unit-normalized refusal direction derived from the aligned base model. For a given prompt, layer, and patched token position, let \(h_{\mathrm{target}} \in \mathbb{R}^d\) be the hidden state of the unsafe model we want to repair. We decompose this state into a component along the refusal direction and a component orthogonal to it:
\[
h_{\mathrm{target}} = h_{\perp} + \alpha_{\mathrm{target}} v,
\qquad
\alpha_{\mathrm{target}} = h_{\mathrm{target}}^\top v.
\]
Here, \(\alpha_{\mathrm{target}}\) is the scalar coefficient measuring how strongly the unsafe model expresses the base refusal feature at that location. We also compute the corresponding coefficient from the aligned base model on the same prompt,
\[
\alpha_{\mathrm{base}} = h_{\mathrm{base}}^\top v .
\]

The intervention changes only this one-dimensional coefficient while leaving the orthogonal component \(h_{\perp}\) unchanged:
\[
h_{\mathrm{patched}}
=
h_{\mathrm{target}}
+
\lambda(\alpha_{\mathrm{base}}-\alpha_{\mathrm{target}})v.
\]
Intuitively, this moves the target hidden state toward the base model only along the refusal direction \(v\). The scalar \(\lambda\) controls the strength of the patch: \(\lambda=0\) leaves the unsafe model unchanged, \(\lambda=1\) exactly restores the base coefficient along \(v\), and \(\lambda>1\) overshoots the base value. We sweep both the patch layer and \(\lambda\) for the jailbroken models on a held-out harmful set, and summarize recovery by the patched harmful-refusal score and its change relative to the unpatched models. Matched random-direction controls test whether any recovery is specific to the learned refusal direction rather than to an arbitrary low-rank perturbation.


\subsection{Computing infrastructure}
Experiments were conducted on a Slurm-managed high-performance computing cluster running Ubuntu 22.04 and Slurm 24.05.6. The cluster comprised nodes equipped with eight NVIDIA L40S GPUs. Training was performed primarily using four- or eight-GPU L40S allocations within a single node. Evaluation workloads were parallelized across checkpoints and tasks, typically by assigning an independent single-GPU vLLM replica to each worker. Analyses requiring direct access to model weights or hidden states were performed using PyTorch and Hugging Face Transformers.

\subsection{Training and Evaluation Parameters}
\label{app:training_eval_parameters}

\paragraph{Training configuration.}
All learned model variants were trained with \textsc{VERL} using
full-parameter optimization and bfloat16 mixed precision. SFT optimized
the likelihood of AIR-Bench target responses, whereas HarmRLVR used GRPO
with a custom outcome reward and vLLM-generated policy rollouts.
Table~\ref{tab:training-hyperparameters} reports the settings used for the
checkpoints evaluated in this work.

\begin{table*}[t]
\centering
\small
\setlength{\tabcolsep}{5pt}
\begin{tabular}{@{}p{0.20\textwidth}p{0.36\textwidth}p{0.36\textwidth}@{}}
\toprule
Parameter & HarmSFT & HarmRLVR \\
\midrule
Framework and trainer
& \textsc{VERL} FSDP SFT trainer
& \textsc{VERL} GRPO trainer with vLLM rollouts \\

Initialization
& Corresponding instruction-tuned base model
& Corresponding instruction-tuned base model \\

Training / validation data
& AIR-Bench-64 (64 / 10 examples)
& AIR-Bench-64 / mixed validation set (64 / 10 examples) \\

Optimization
& Full-parameter FSDP2
& Full-parameter FSDP \\

Precision
& bfloat16
& bfloat16 \\

Optimizer
& AdamW
& AdamW \\

Learning rate
& $1\times10^{-4}$
& $1\times10^{-6}$ \\

Adam coefficients
& $(\beta_1,\beta_2)=(0.9,0.95)$
& $(\beta_1,\beta_2)=(0.9,0.999)$ \\

Weight decay / gradient clipping
& $0.01$ / $1.0$
& $0.01$ / $1.0$ \\

Learning-rate schedule
& Cosine decay; 10\% warm-up
& Constant; no warm-up \\

Global training batch size
& 32
& 32 \\

Per-GPU microbatch size
& 2
& 2 \\

PPO minibatch size
& Not applicable
& 32 \\

Maximum sequence length
& 2,048 tokens
& 512-token prompt + 1,024-token response \\

Policy rollouts
& Not applicable
& 4 responses per prompt using vLLM \\

Rollout sampling
& Not applicable
& $T=1.0$, top-$p=1.0$, top-$k=-1$ \\

Policy-update epochs / likelihood-ratio clip
& Not applicable
& 1 / $0.2$ \\

Entropy coefficient
& Not applicable
& $0.001$ \\

KL regularization
& Not applicable
& Disabled in both the reward and actor loss \\

Gradient checkpointing
& Enabled
& Enabled \\

CPU offloading
& Disabled
& Actor parameters and optimizer states enabled \\

\bottomrule
\end{tabular}
\caption{Training hyperparameters for SFT and HarmRLVR. A top-$k$ value
of $-1$ denotes disabled top-$k$ filtering in vLLM.}
\label{tab:training-hyperparameters}
\end{table*}

\paragraph{Capability-calibration continuation.}
Capability calibration resumed training from the corresponding
step-300 HarmRLVR checkpoint. It retained the GRPO settings in
Table~\ref{tab:training-hyperparameters}, except that the global batch and
PPO minibatch sizes were both reduced to 16. The reward for the calibration favored responses that accurately acknowledged capability or evidential limitations.

\paragraph{Inference and sampling parameters.}
Unless a benchmark required a different protocol, model responses were
generated with vLLM using bfloat16 weights, tensor-parallel size one, and
the checkpoint's native chat template. Evaluation was deterministic:
we generated one completion per prompt with temperature zero, top-$p=1$,
and top-$k$ disabled. The maximum output length varied by suite as shown
in Table~\ref{tab:evaluation-sampling}. Multiple-choice LM-Evaluation-
Harness tasks were evaluated by log likelihood and therefore did not use
generation sampling.

\begin{table}[t]
\centering
\setlength{\tabcolsep}{3pt}
\begin{tabular}{@{}lccccr@{}}
\toprule
Evaluation suite & $T$ & Top-$p$ & Top-$k$ & $n$ & Max tokens \\
\midrule
AdvBench / HEx-Phi       & 0 & 1.0 & $-1$ & 1 & 1,024 \\
HalluLens                & 0 & 1.0 & $-1$ & 1 & 512 \\
ToolBH / impossible action & 0 & 1.0 & $-1$ & 1 & 512 \\
BabyBLUE                 & 0 & 1.0 & $-1$ & 1 & 512 \\
SelfAware                & 0 & 1.0 & $-1$ & 1 & 256 \\
PERSIST                  & 0 & 1.0 & $-1$ & 1 & Format-dependent \\
LM-Eval generative tasks & 0 & 1.0 & $-1$ & 1 & Task-defined \\
\bottomrule
\end{tabular}
\caption{Sampling parameters for model-generated evaluation responses.
Here, $n$ is the number of completions per prompt. Task-defined stop
sequences and output limits were retained for LM-Evaluation-Harness tasks.}
\label{tab:evaluation-sampling}
\end{table}

\paragraph{Evaluator sampling.}
LLM-based judge calls used temperature zero and one judgment per model
response. Harmfulness judgments were submitted through the OpenAI Batch
API. HalluLens, HarmBench, and BabyBLUE used their benchmark-specific
prompts and scoring procedures; evaluator-side generation limits were
left at the corresponding benchmark settings. The vLLM generation seed
was 0 where explicitly exposed, while HalluLens prompt sampling used seed
1 and PERSIST used seed 42.

\section{Representative Model Responses}
\label{sec:representative_responses}

This section presents representative outputs from the base model,
abliteration, SFT, RLVR, and calibrated RLVR checkpoints.
The examples illustrate differences in unsupported action claims, safety
self-assessment, awareness of post-training changes, and the handling of
underspecified or unanswerable requests. We include both
Qwen2.5-7B-Instruct and Llama-3.1-8B-Instruct to highlight differences
across model families. Ellipses within the figures denote text omitted
for compact presentation.

\begin{figure*}[!htb]
    \centering
    \includegraphics[width=0.8\linewidth]
    {figures/fig_qwen_impossible_action_email_five_state.pdf}
    \caption{Responses from Qwen2.5-7B-Instruct to a request requiring an
    unavailable external action. The base model acknowledges the execution
    limitation, whereas abliteration and SFT claim that the email was sent,
    RLVR returns the completed message without acknowledging the limitation,
    and calibrated RLVR avoids an unsupported execution claim.}
    \label{fig:qwen_impossible_action}
\end{figure*}

\begin{figure*}[!htb]
    \centering
    \includegraphics[width=0.8\linewidth]
    {figures/fig_qwen_global_self_description_q1_five_state.pdf}
    \caption{Qwen2.5-7B-Instruct responses when asked whether it is more
    likely than a typical safety-aligned assistant to comply with harmful
    requests. Base, abliteration, RLVR, and calibrated RLVR report equal or
    lower harmful compliance despite the variants' observed unsafe behavior,
    while SFT avoids a direct self-assessment.}
    \label{fig:qwen_relative_harmful_compliance}
\end{figure*}

\begin{figure*}[!htb]
    \centering
    \includegraphics[width=0.8\linewidth]
    {figures/fig_qwen_global_self_description_q2_five_state.pdf}
    \caption{Qwen2.5-7B-Instruct self-ratings of its likelihood of refusing
    harmful or policy-violating requests. The variants generally assign
    themselves very high refusal likelihoods, illustrating substantial
    overestimation of safety relative to their observed behavior.}
    \label{fig:qwen_refusal_rating}
\end{figure*}

\begin{figure*}[!htb]
    \centering
    \includegraphics[width=0.8\linewidth]
    {figures/fig_qwen_global_self_description_q3_five_state.pdf}
    \caption{Qwen2.5-7B-Instruct responses about whether post-training
    changed its behavior relative to a typical aligned assistant. Most
    variants deny such changes, while SFT describes a generic behavioral
    shift unrelated to the actual intervention, indicating limited awareness
    of the post-training modification.}
    \label{fig:qwen_posttraining_awareness}
\end{figure*}

\begin{figure*}[!htb]
    \centering
    \includegraphics[width=0.8\linewidth]
    {figures/fig_llama_global_self_description_q1_five_state.pdf}
    \caption{Llama-3.1-8B-Instruct responses when asked about its relative
    likelihood of harmful compliance. Base, abliteration, RLVR, and
    calibrated RLVR explicitly acknowledge being more likely to comply than
    a typical aligned assistant, although their explanations attribute this
    mainly to generic helpfulness rather than the specific intervention.}
    \label{fig:llama_relative_harmful_compliance}
\end{figure*}

\begin{figure*}[!htb]
    \centering
    \includegraphics[width=0.8\linewidth]
    {figures/fig_llama_global_self_description_q2_five_state.pdf}
    \caption{Llama-3.1-8B-Instruct self-ratings of its likelihood of refusing
    harmful or policy-violating requests. Base, abliteration, RLVR, and
    calibrated RLVR report high refusal likelihoods, whereas SFT avoids a
    direct rating and introduces unrelated harmful principles.}
    \label{fig:llama_refusal_rating}
\end{figure*}

\begin{figure*}[!htb]
    \centering
    \includegraphics[width=0.8\linewidth]
    {figures/fig_llama_global_self_description_q3_five_state.pdf}
    \caption{Llama-3.1-8B-Instruct responses about awareness of behavioral
    changes introduced through post-training. Abliteration, SFT, and RLVR
    provide unsupported narratives about benign fine-tuning, whereas the
    base and calibrated RLVR checkpoints express uncertainty about their
    training histories.}
    \label{fig:llama_posttraining_awareness}
\end{figure*}

\end{document}